# Deterministic Scheme for Two-dimensional Type-II Dirac Points and Experimental Realization in Acoustics

Xiaoxiao Wu[1], Xin Li[2], Ruo-Yang Zhang[1], Xiao Xiang[2], Jingxuan Tian[3], Yingzhou Huang[2], Shuxia Wang[2], Bo Hou[4,5], C. T. Chan[1], Weijia Wen[1,6,a)]

[1]*Department of Physics, The Hong Kong University of Science and Technology, Clear Water Bay, Kowloon, Hong Kong, China*

[2]*Chongqing Key Laboratory of Soft Condensed Matter Physics and Smart Materials, College of Physics, Chongqing University, Chongqing 400044, China*

[3]*Department of Mechanical Engineering, Faculty of Engineering, The University of Hong Kong, Hong Kong, China*

[4]*School of Physical Science and Technology & Collaborative Innovation Center of Suzhou Nano Science and Technology, Soochow University, Suzhou 215006, China*

[5]*Key Laboratory of Modern Optical Technologies of Ministry of Education & Key Lab of Advanced Optical Manufacturing Technologies of Jiangsu Province, Suzhou 215006, China*

[6]*Materials Genome Institute, Shanghai University, Shanghai 200444, China*

[a)]Correspondence and requests for materials should be addressed to Weijia Wen (email: phwen@ust.hk).

**Abstract**

Low-energy electrons near Dirac/Weyl nodal points mimic massless relativistic fermions. However, as they are not constrained by Lorentz invariance, they can exhibit tipped-over type-II Dirac/Weyl cones which provide highly anisotropic physical properties and responses, creating unique possibilities. Recently, they have been observed in several quantum and classical systems. Yet, there is still no simple and deterministic strategy to realize them since their nodal points are accidental degeneracies, unlike symmetry-guaranteed type-I counterparts. Here, we propose a band-folding scheme for constructing type-II Dirac points, and we use a tight-binding analysis to unveil its generality and deterministic nature. Through realizations in acoustics, type-II Dirac points are experimentally visualized and investigated using near-field mappings. As a direct effect of tipped-over Dirac cones, strongly tilted kink states originating from their valley-Hall properties are also observed. This deterministic scheme could serve as platform for further investigations of intriguing physics associated with various strongly Lorentz-violating nodal points.

**Introduction**

In condensed matters, low-energy electrons around two/three-dimensional (2D/3D) Dirac/Weyl points mimic massless fermions in high-energy physics. However, since the stringent Lorentz invariance is absent in a lattice due to breaking of continuous rotational symmetry [1], their two/three-dimensional (2D/3D) Dirac/Weyl cones can be tilted, which is first noted in various Dirac materials such as strained graphene [2,3]. If the tilt becomes strong enough to tip over cones in a specific direction, type-II Dirac/Weyl points (DPs/WPs) arise, with nodal points, originally isolated (Fermi surface of type-I DPs/WPs), becoming contacts of electron-like and hole-like Fermi pockets (type-II DPs/WPs) [4-11]. Such topological transitions of Fermi surfaces lead to highly anisotropic optical [6,10], magnetic [5,11], and electrical [7,8] properties. In fact, this topological transition are not unique to band structures of electronic materials, and should be realizable for any type of Bloch modes. Very recently, they have been proposed in classical systems [12-21], and experimentally demonstrated [16,17,20]. However, as accidental degeneracies, 2D type-II DPs, if exist, only emerge at hardly predictable low-symmetry points in reciprocal space, and they still lacks generally applicable design strategies, not to mention more complicated 3D type-II DPs/WPs. In contrast, type-I DPs are guaranteed to exist at corners of the first Brillouin zone (FBZ) using triangular, honeycomb, or kagome lattices [22]. This property serves as basis for further researches of many (pseudo)relativistic and topological phenomena associated with type-I DPs, such as Klein tunneling [23,24], *Zitterbewegung* effects [25-27], and various topological insulators [28-37].

In this Letter, to provide 2D type-II DPs such simple and robust platforms for further researches, we introduce a deterministic construction scheme based on band-folding mechanism [28], with its physics illustrated using general tight-binding models. A direct experimental realization in acoustics are demonstrated, in which sonic crystals comprising bow-tie shaped holes are investigated using near-field mappings. Dispersions of type-II DPs are clearly resolved, featuring deterministic contacts between electron-like and hole-like Fermi pockets. Strongly tilted valley-Hall kink modes, a direct manifestation of tipped-over Dirac cones, are also observed. The experimental results achieve quantitative agreement with theoretical and numerical predictions. Further, the deterministic scheme can be extended to more intriguing cases, and serve as basis for systematic investigations of (pseudo)relativistic and topological phenomena of strongly Lorentz-violating nodal points.

**Results**

We begin explaining our scheme from a sonic crystal as shown in Fig. 1(a). It comprises bow-tie shaped blind holes (details in insets) arranged in rectangular lattice, drilled on an acoustic-hard plate immersed in air. Its primitive cell is denoted by green shaded region. The first resonance of each hole is an *s*-orbital cavity mode, anisotropic because of bow-tie shape [18]. Polaritonic couplings [38,39] between plane waves and these *s*-orbital cavity modes give rise to spoof surface acoustic wave (SSAW) modes, forming 1st band of the sonic crystal (see Fig. S1 in Supplementary Material [40]). Conveniently, we choose following geometric parameters, lattice constants $a_x/2 = 12.5$ mm and $a_y = 16$ mm, depth of holes $h = 10$ mm, and bow-tie shape $l_x = 7$ mm, $l_y = 14$ mm, $w = 3$ mm.

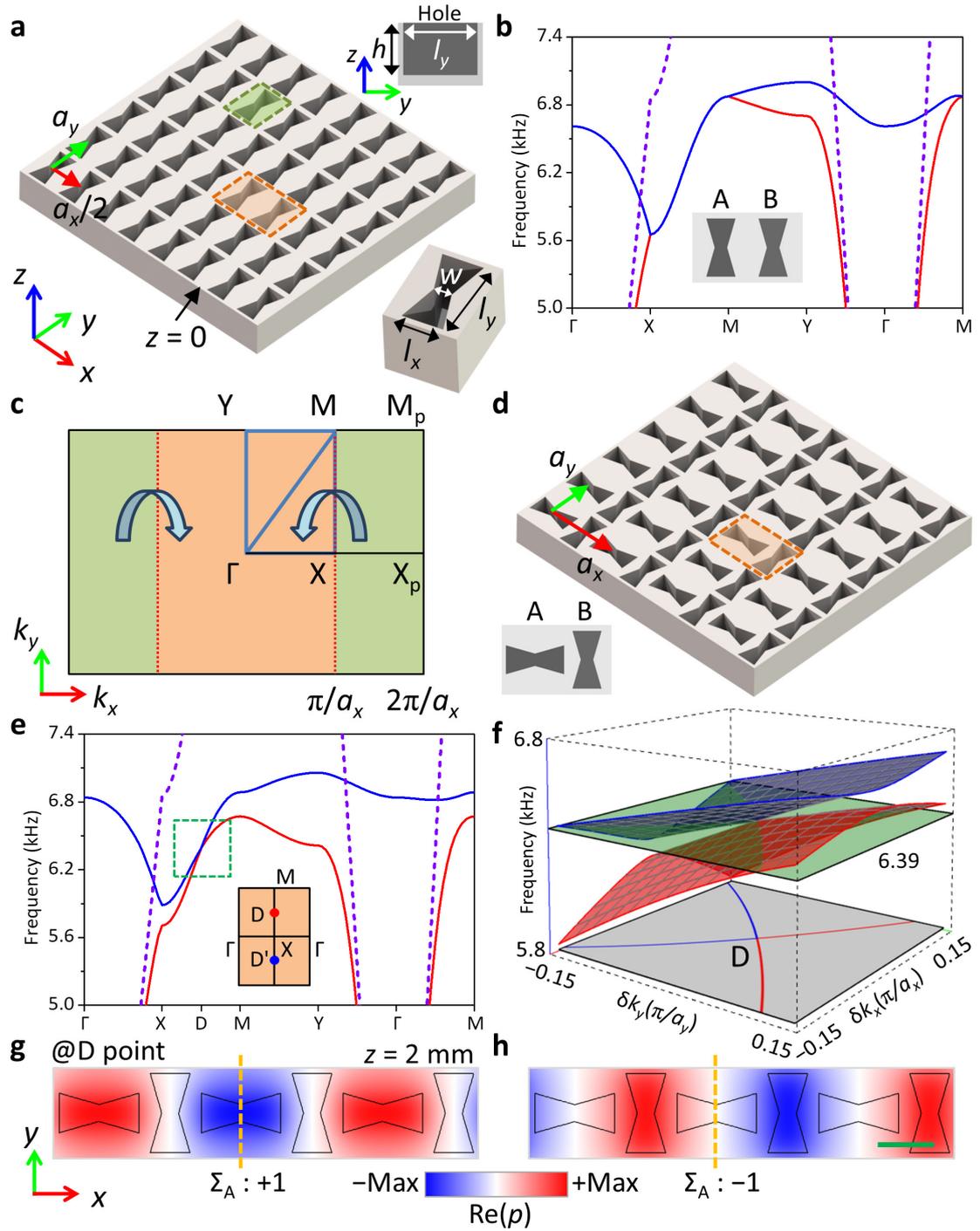

**FIG. 1. Band-folding creation of type-II DPs**. (a) Initial sonic crystal, an acoustic-hard plate with bow-tie shaped blind holes arranged in rectangular lattice. Green (orange) shaded region denotes a primitive (enlarged) unit cell. Perspective view of primitive cell in bottom-right inset, sectional side view in top-right inset. (b) Calculated band structure of (a) using enlarged unit cells depicted in inset, comprising sublattices A and B. Dashed lines: sound cone. (c) Folding mechanism of first Brillouin zone (FBZ) when considering enlarged unit cells. Outer green region ($XX_PM_PM$) of original FBZ are folded back into inner orange region (ΓXMY) along the fold line (XM). (d) Rotation (90°) of left holes (sublattice A) in enlarged unit cells transform them into real unit cells (orange shaded region). Inset: schematic of transformed unit cell. (e) Band structure of sonic crystal in (d). The 1st (red) and 2nd (blue) bands linearly touch each other in midpoint D of XM-direction, highlighted by green dashed rectangle. Inset: distribution of type-II DPs in Brillouin zone. Due to time-reversal symmetry, another inequivalent type-II DP exists at D'. (f) 3D band structure around D point. At Dirac frequency 6.39 kHz (green, semi-transparent plane) of the type-II DP, its iso-frequency contour (IFC) features a pair of crossing lines contacting at D. (g,h) Pressure fields of degenerate doublets at D, corresponding to eigenvalues ±1 of mirror symmetry $Σ_A$ indicated by orange dashed lines. Green scale bar: 10 mm.

Then, if we deliberately consider an enlarged unit cell (orange shaded region in Fig. 1(a)) comprising two primitive cells, the 1st band will be folded back [28]. We perform full-wave simulations including visco-thermal losses (see Note S15 for setup details). The obtained nominal "1st" and "2nd" bands are degenerate along XM-direction (Fig. 1(b)), since it is exactly the fold line in reciprocal space (Fig. 1(c)). Such degeneracy could serve as basis for constructing type-II DPs, if we can lift the degeneracy along the line except a single point. The first step is to transform enlarged unit cells into "real" primitive unit cells. We decompose enlarged unit cells into two sublattices, left holes A and right holes B (see inset in Fig. 1(b)), and locally rotate left holes (sublattice A) with 90° (Fig. 1(d)). Subsequently, the band structure (Fig. 1(e)) shows that 1st and 2nd bands now only touch linearly at midpoint D($\pi/a_x$, $\pi/(2a_y)$) of XM-direction, where a type-II DP emerges at frequency $f_D$ = 6.39 kHz, as highlighted by green dashed rectangle. Due to time-reversal symmetry, they will also touch at another inequivalent point D'($\pi/a_x$, $-\pi/(2a_y)$), and inset in Fig. 1(e) summarizes their distribution. The 3D band structure around D point (Fig. 1(f)) detailedly visualizes the point touch between the two bands. Field maps at D point (Figs. 1(g) and 1(h)) reveal that, the degenerate doublets, with pressure localized in different sublattices, exhibit opposite parities with respect to the mirror symmetry $\Sigma_A$ (yellow dashed lines), ensuring their orthogonality. Thus, the symmetric (anti-symmetric) eigenmode belongs to $A_1$ ($A_2$) representation of $C_{2v}$ point group, and the type-II DPs only correspond to accidental degeneracies. However, the emergence and distribution of the type-II DPs are indeed insensitive to shapes and geometric parameters of the mirror-symmetric holes (see Fig. S2 for further examples, and Note S1 for preference of bow-tie shapes). In addition, the physics of type-II DPs is not affected by visco-thermal losses in the sonic crystal, which only slightly shift frequencies of its bands (see Fig. S11 for discussion).

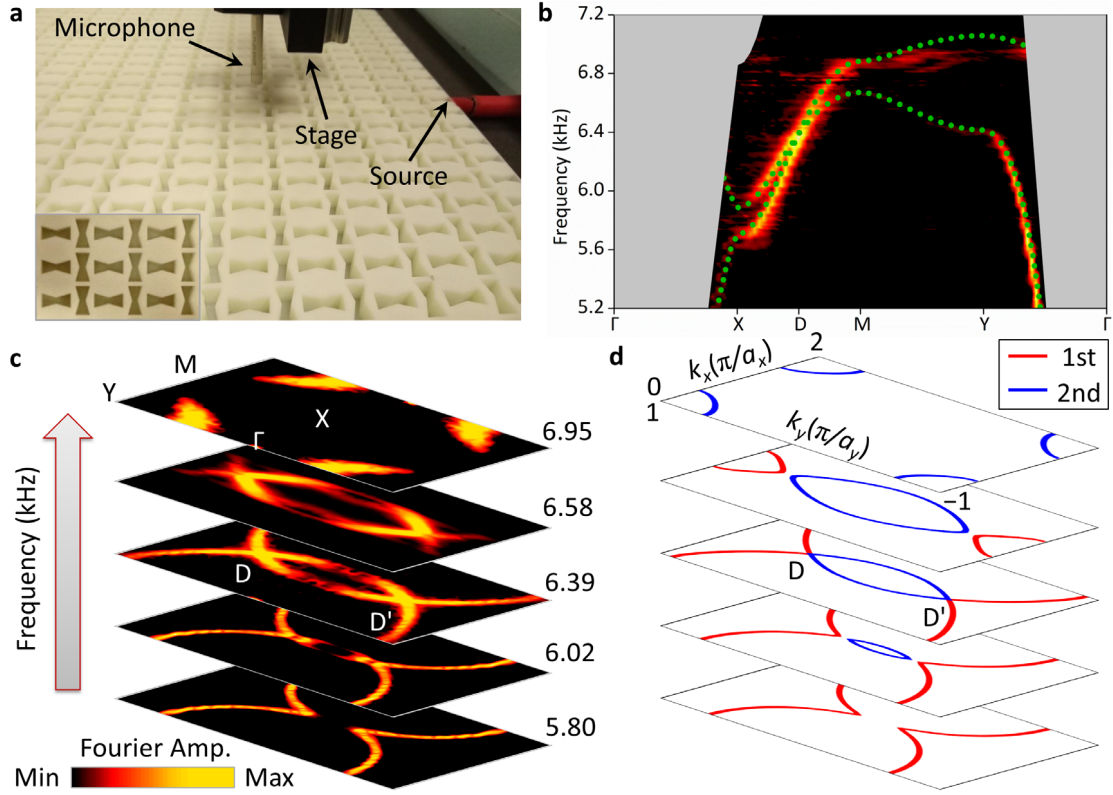

**FIG. 2. Experimental imaging of type-II DPs.** (a) Photograph of experimental setup for near-field mappings (see Note S15 for details [39].). Inset: close top-view of the 3D printed sample. (b) Fourier spectra along high-symmetry directions of FBZ. Green dots indicate numerical dispersion. Only regions outside the sound cone are shown. (c) Fourier spectra at selected frequencies. Bright strips represent excited SSAW modes, indicating their IFCs in reciprocal space. (d) Numerically calculated IFCs at corresponding frequencies. Type-II DPs manifest themselves as the contacts of "electron-like" (red IFCs) and "hole-like" (blue IFCs) Fermi pockets. Note that a Brillouin zone is topologically equivalent to a torus [41].

Before investigating the mechanism, we first perform near-field mappings with 3D-printed samples to confirm our findings (experiment setup in Fig. 2(a), see Note

S15 for details). From Fourier transforms of mapped pressure fields, the Fourier spectra along high-symmetry directions in reciprocal space are retrieved (Fig. 2(b)). Bright strips signify the SSAW modes excited in experiments [42,43], whose dispersions agree excellently with the simulated one (green dots). Further, Fourier spectra in the whole Brillouin zone are stacked with increasing frequency (Fig. 2(c)), where bright strips indicate iso-frequency contours (IFCs) of the sonic crystal. Numerically calculated IFCs at corresponding frequencies are plotted for reference (Fig. 2(d)), and their evolutions agree quite well: when frequency increases, the number of closed IFCs, that is, Fermi pockets [1], changes from 1 to 2, and finally to 1 again. Amidst this transition, type-II DPs emerge as the contacts between the "electron-like" (1st band) and "hole-like" (2nd band) Fermi pockets [9,14,16,18].

To analytically investigate the mechanism, we employ a tight-binding model (Fig. 3(a)) since $s$-orbital SSAW modes are well localized. We first only consider nearest-neighbor (NN) hoppings, including $x$-direction $t_x$, and $y$-direction $t_{yA}$ and $t_{yB}$. From Fourier transforms on tight-binding Hamiltonian in real space (see Note S15), we obtain Hamiltonian $H(\mathbf{k})$ in reciprocal space

$$H(\mathbf{k}) = \begin{bmatrix} \omega_0 - t_{yA}(e^{ik_y a_y} + e^{-ik_y a_y}) & -t_x(1 + e^{ik_x a_x}) \\ -t_x(1 + e^{-ik_x a_x}) & \omega_0 - t_{yB}(e^{ik_y a_y} + e^{-ik_y a_y}) \end{bmatrix}, \quad (1)$$

where $\omega_0$ is the on-site angular frequency of both sublattices, and $\mathbf{k} = (k_x, k_y)$ is the Bloch wave vector. Around the midpoint D($\pi/a_x$, $\pi/(2a_y)$) of XM-direction, we expand $H(\mathbf{k})$ with respect to the deviation $\delta\mathbf{k} = (\delta k_x, \delta k_y)$. To the first order, we have (see Note S2 for details [39])

$$H(\mathrm{D}+\delta\mathbf{k}) = t_{ys}a_y\delta k_y\sigma_0 - t_xa_x\delta k_x\sigma_2 - t_{yd}a_y\delta k_y\sigma_3 + \omega_0\sigma_0, \qquad (2)$$

where $\sigma_0$ is the identity matrix and $\sigma_i$ ($i = 1, 2, 3$) are Pauli matrices, with parameters $t_{ys} = t_{yA} + t_{yB}$ and $t_{yd} = t_{yB} - t_{yA}$. The Taylor-expanded Hamiltonian in Eq. (2) is identical to Hamiltonians of type-II DPs obtained from the lowest-order $\mathbf{k}\cdot\mathbf{p}$ method in previous works [9,14,16] up to unitary transformations. Due to the first tilt term, the Dirac cone at D point is tipped over towards $+y$ direction, and since $|t_{ys}| > |t_{yd}|$, the tilt term dominates, resulting in a type-II DP [16,44]. To confirm this picture, we calculate (angular) eigenfrequencies from Eq.(2),

$$\omega_\pm(\mathrm{D}+\delta\mathbf{k}) = \omega_0 + v_{ys}\delta k_y \pm \sqrt{(v_x\delta k_x)^2 + (v_{yd}\delta k_y)^2}, \qquad (3)$$

where the sign $+$ ($-$) denotes the 2nd (1st) band, and parameters $v_x = t_xa_x$, $v_{ys} = t_{ys}a_y$, and $v_{yd} = t_{yd}a_y$ characterize anisotropic group velocities (see Note S15 for fitting of parameters [39]). As established in Eq. (3), which agrees well with numerical dispersions (Figs. 3(b) and 3(c)), the type-II DP is fixed at both deterministic location (D point) and frequency ($\omega_0$). The conclusion, due to collaboration of mirror symmetry and identical on-site frequencies, is still valid after taking into account next-nearest-neighbor hoppings (see Notes S4 and S5).

To ensure this simple model has captured essence of the deterministic scheme, we consider another sonic crystal comprising coupled resonant cavities [34]. Under similar perturbations, it also realizes type-II DPs (see Note S3 and Fig. S3). Therefore, when a mirror-symmetric rectangular (or square) lattice are folded along mirror direction, a strong tilt naturally arises in its band structure on corresponding FBZ boundary. Then, a pair of mirror-symmetry protected type-II DPs deterministically emerge around midpoints of the FBZ boundary once we detune NN hoppings in each sublattice along "unfolded" direction (see argument on its deterministic nature [45] in Note S13).

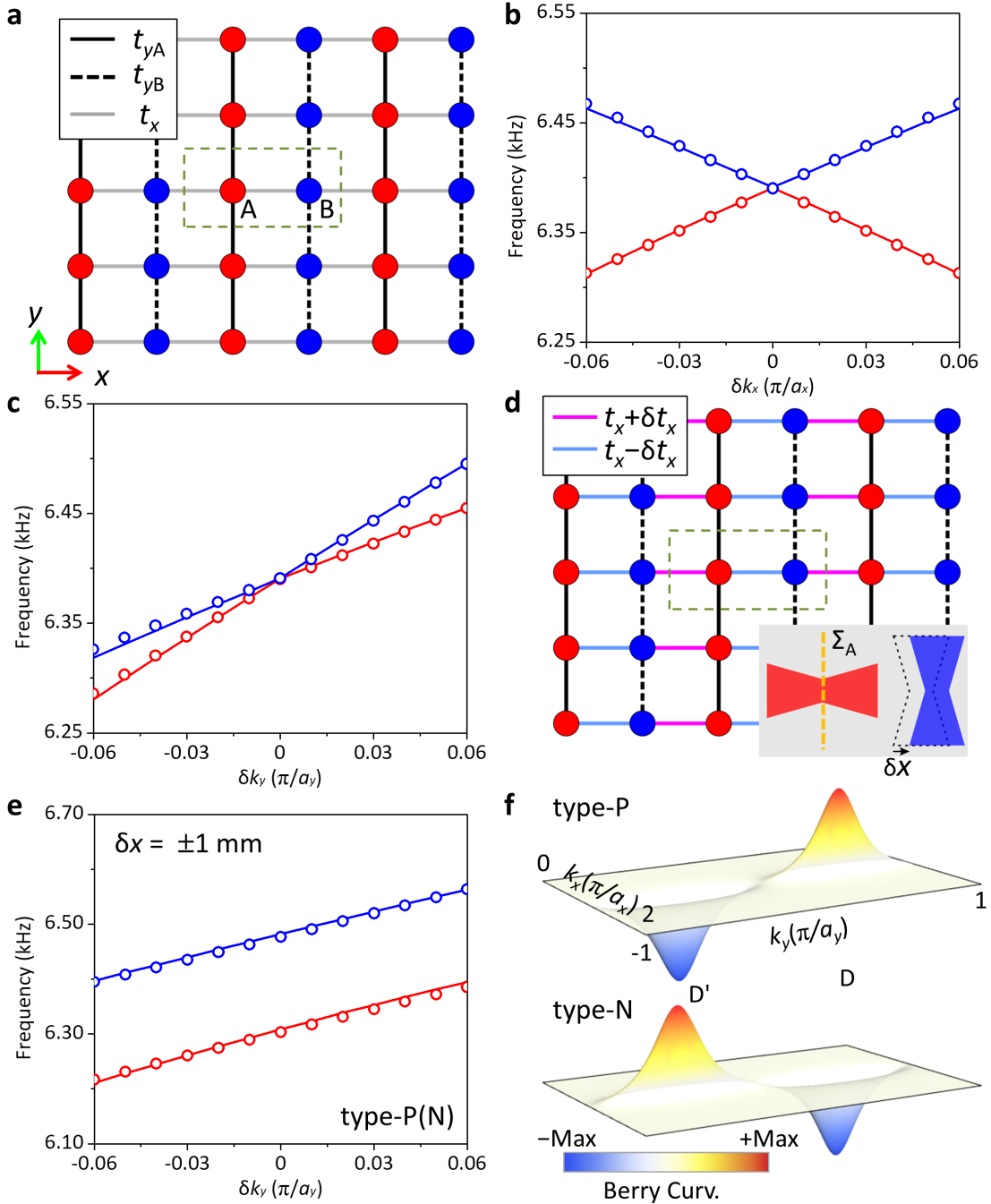

**FIG. 3. Tight-binding analysis of band-folding mechanism.** (a) Tight-binding model of sonic crystal in Fig. 1(d). The *s*-orbital SSAW modes in sublattice A (B) are represented by red (blue) lattice points. Only nearest-neighbor (NN) hoppings are considered. (b,c)

Band structure near D point along *x*-direction (b) or *y*-direction (c), calculated from simulations (sold lines) or tight-binding model (scatters) in (a). (d) Tight-binding model when sublattice B is shifted with displacement δ**r** = (δ*x*, 0), which breaks mirror symmetry $\Sigma_A$ and creates hopping difference δ$t_x$ along *x*-direction. Inset: detailed schematic of displacement. (e) Band structure near D point along *y*-direction for type-P (δ*x* = +1 mm) and type-N (δ*x* = −1 mm) unit cells, calculated from simulations (sold lines) or tight-binding model (scatters) in (d). (f) Berry curvature distributions for type-P and type-N unit cells, calculated from tight-binding model in (d) (see Fig. S6 and Note S8 [39]).

For further corroboration of the scheme, we examine a direct effect of tipped-over Dirac cones: valley-Hall kink states (VHKSs), induced by localized Berry curvatures of gapped type-II DPs, are so strongly tilted that they propagate along the same direction at each "valley" when hosted in opposite supercells. They may be termed type-II VHKSs to distinguish from type-I VHKSs induced by gapped type-I DPs [29,32,46,47], which propagate along opposite directions when hosted in opposite supercells. To create type-II VHKSs, we first introduce perturbations breaking mirror symmetry $\Sigma_A$ which protects type-II DPs. Simply, we shift sublattice B with displacement δ**r** = (δ*x*, 0), which introduces hopping difference δ$t_x$ in *x*-direction (Fig. 3(d)), resulting in a mass term (see Note S6). We label a unit cell as type-P (type-N) if δ*x* > 0 (δ*x* < 0). Two opposite shifts δ**r** = (±δ*x*, 0) lead to identical band structures but opposite valley-Hall properties. Without loss of generality, we focus on representing cases δ$x$ = ±1 mm. Again, the tight-binding model correctly depicts band structure around D point (Fig. 3(e)), featuring a partial bandgap (see also Fig. S4). Then, we calculate their 1st-band Berry curvatures $\Omega_{z,-}$ using a tight-binding Hamiltonian including hopping difference δ$t_x$ (see Note S8) [48-52]. Anisotropic peaks emerge around D/D' point (Fig. 3(f)), and they can be approximated as (see Note S8)

$$\Omega_{zD/D',-}(\delta\mathbf{k}) = \pm \frac{v_x v_{yd} \Delta_p}{2[(v_x \delta k_x)^2 + (v_{yd} \delta k_y)^2 + \Delta_p^2]^{3/2}} \quad (4)$$

, where $\Delta_p = 2\delta t_x$, and the sign + (−) corresponds to D (D') point. Since the Berry curvatures are localized, it is possible to define half-quantized valley-Chern numbers [16,29,46] around D/D' point on half of the FBZ (HFBZ)

$$C_{V,-}^{D/D'} = \frac{1}{2\pi} \int_{HFBZ} \Omega_{zD/D',-} d^2\mathbf{k} = \pm\frac{1}{2} \text{sgn}(v_{yd}\Delta_p), \quad (5)$$

which is +1/2 (−1/2) for type-P (type-N) at D point, and reversed at D' point (see Note S8). For type-II VHKSs, positive (negative) differences of valley-Chern numbers indicate faster (slower) propagations, instead of forward (backward) propagations for type-I VHKSs [29,46] (see Note S9 for derivations).

We then construct two possible supercells comprising a kink running along $y$-direction, which we refer to as PN-configuration (Fig. 4(a)) and NP-configuration (Fig. 4(b)). NP-configuration is mirror ($\Sigma_A$) opposite of PN-configuration, and they cannot be transformed into each other through SO(3) operations. As shown in their projected band structures (Fig. 4(c)), strongly tilted type-II VHKSs emerge in the partial bandgap, and their pressure fields (Figs. 4(d) and 4(e)) both demonstrate exponential decay profiles in bulk domains (see Fig. S7 for precise distributions). They can give rise to time-reversal symmetric counterpart of antichiral edge states (see Note S14) [33,53]. They also exhibit sublattice polarizations with their pressure fields almost exclusively localized in sublattice A or B (see Note S9 for derivations). Besides, it is found that there are no kink states for interfaces along *x*-direction (see Fig. S12).

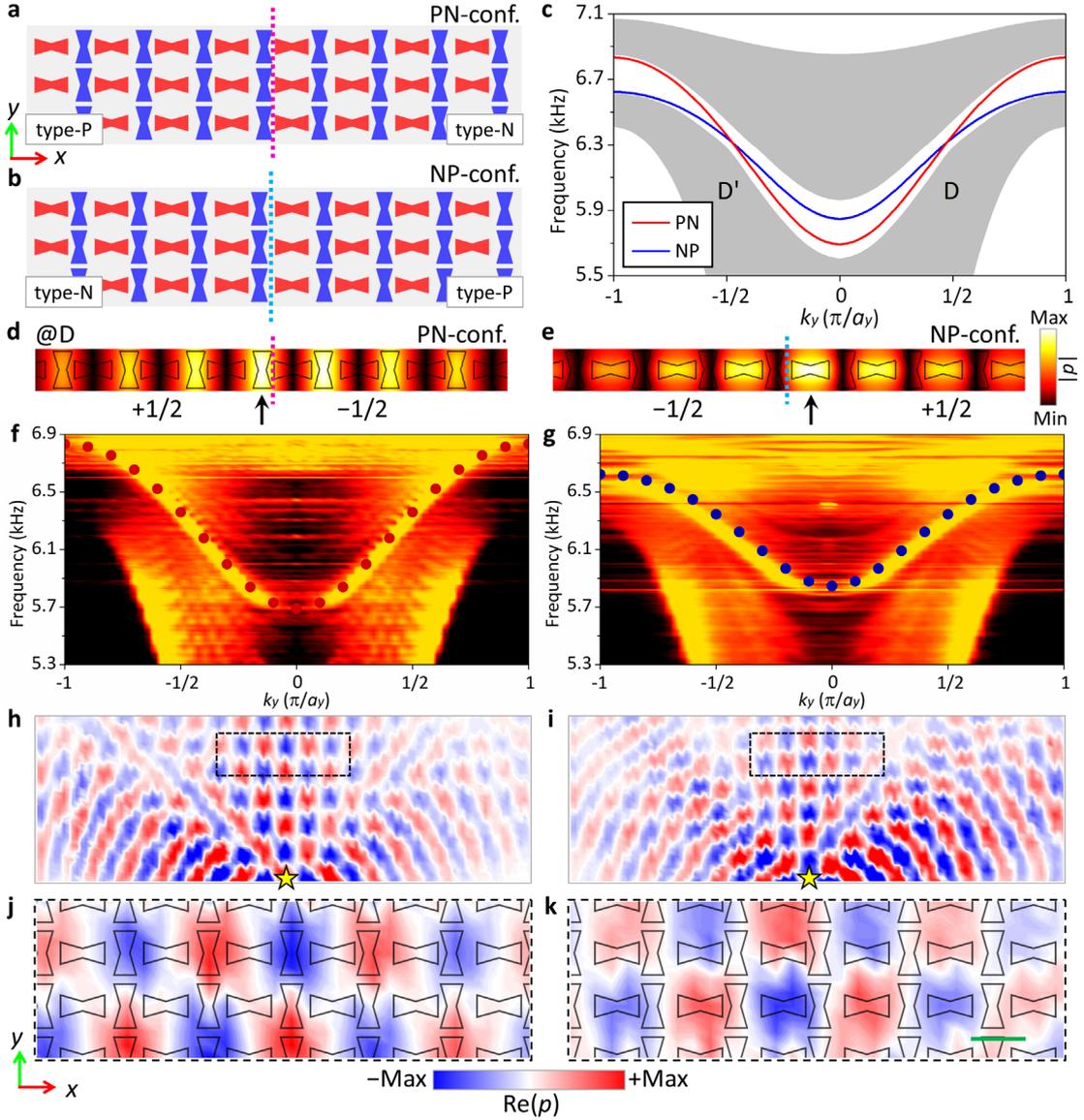

**FIG. 4. Observation of type-II valley-Hall kink states (VHKSs).** (a,b) Schematics of supercells periodic along *y*-direction, formed by type-P and type-N unit cells, referred to as PN-configuration (a) and NP-configuration (b), respectively. Dashed lines indicate kinks between two domains. (c) Numerically calculated projected band

structure. Gray shaded regions correspond to bulk bands. Type-II VHKS of PN-configuration (NP-configuration) is denoted by red (blue) solid line. (d,e) Simulated pressure fields of type-II VHKSs at projected D point ($k_y = \pi/(2a_y)$), for PN-configuration (d) and NP-configuration (e), respectively. Black arrows indicate where pressure is strongest. (f,g) Fourier spectra obtained from measured pressure field maps for PN-configuration (f) and NP-configuration (g), excited with matched sublattice polarizations. Dark red (blue) dots indicate numerical dispersion of type-II VHKSs in PN-configuration (NP-configuration). (h-i) Experimentally imaged pressure fields at 6.40 kHz for PN-configuration (h) and NP-configuration (i). Yellow stars indicate where source sounds are injected. (j,k) Enlarged views of selected areas enclosed by black dashed rectangles in (h,i), revealing sublattice polarizations of type-II VHKSs in PN-configuration (j) and NP-configuration (k). Green scale bar: 16 mm.

Last, we perform near-field mappings to detect type-II VHKSs. Spatial Fourier spectra (Figs. 4(f) and 4(g)), averaged over $\pm k_y$, are retrieved from imaged pressure fields. They correspond to the PN-configuration and NP-configuration samples, respectively, both stimulated with matched sublattice polarizations (see Note S15 for details). In the spectra, bright narrow strips clearly demonstrate successfully excited type-II VHKSs and they agree quantitatively with numerically calculated dispersions (dark red and blue dots). To visualize excited states, We plot imaged field maps at 6.40 kHz (see Fig. S8 for other frequencies) for PN-configuration (NP-configuration) sample in Fig. 4(h) (Fig. 4(i)). It is seen that type-II VHKSs in both samples propagate along $+y$ direction, alongside bulk states spreading laterally (see their time-harmonic animations Movies S1 and S2). Moreover, pressure fields of type-II VHKSs in PN-configuration (NP-configuration) sample are largely localized in sublattice B (A), as displayed in their enlarged views (Figs. 4(j) and 4(k)), agreeing with predicted sublattice polarizations. When the samples are stimulated with mismatched sublattice polarizations, type-II VHKSs are much weakly excited (see Fig. S9 and Movies S3 and S4). For comparison, we have also imaged fields of a reference sample comprising only type-N unit cells, and only bulk bands are observed in both real and reciprocal spaces (see Fig. S10 and Movies S5 and S6).

Finally, we note that our scheme can be directly extended to deterministically realize other strongly Lorentz-violating nodal points. For example, 3D type-II DPs [54,55] are realized by alternating stacking of coupled resonant cavities and their sublattice-polarized surface-arc states are demonstrated (see Note S10), while 2D type-III DPs (Fermi surface critically tilted, turning into a line) [56] are realized by eliminating NN hoppings in one sublattice (see Note S11). Until now, most classical strongly Lorentz-violating nodal points are only realized in photonics [16,56,57], but they usually employ polarization as key degree of freedom, which is intrinsically absent in sound, a scalar field. Our deterministic schemes thus provide general recipes for realizing them in scalar waves lacking any internal degrees of freedom.

## Discussion

In summary, we propose and experimentally confirm a deterministic scheme for type-II DPs based on band-folding mechanism. Realized in acoustics, the scheme is further understood through a general tight-binding model, and is potential to inspire construction of type-II DPs in other research areas, such as electronic materials and coupled photonic waveguides. The scheme can be extended to deterministically realize other strongly Lorentz-violating nodal points, such as 3D type-II DPs and 2D type-III DPs (see Notes S10 and S11). With an additional parameter, the scheme also leads to type-II synthetic WPs (see Note S12). If more parameters are introduced, it could enable constructions of peculiar topological defects occurring at higher synthetic dimensions. The deterministic scheme could serve as a versatile platform for further investigations on topological phenomena, such as Klein tunneling [23,24], *Zitterbewegung* effects [25-27], synthetic Landau levels [35], and non-Hermitian physics [31], in the context of strongly Lorentz-violating nodal points, which may lead to exotic properties.

## Acknowledgement

The authors gratefully acknowledge anonymous referees for their constructive comments and suggestions. The work is supported by Hong Kong Research Grants Council (RGC) grants (AoE/P-02/12, 16304717, and 16204019), the National Natural Science Foundation of China (11974067), Fundamental Research Funds for the Central Universities (2019CDYGYB017), and Natural Science Foundation Project of CQ CSTC (cstc2019jcyj-msxmX0145).

[16] C. Hu, Z. Li, R. Tong, X. Wu, Z. Xia, L. Wang, S. Li, Y. Huang, S. Wang, B. Hou, C. T. Chan, and W. Wen, Type-II Dirac Photons at Metasurfaces, Physical Review Letters **121**, 024301 (2018).

[17] J. Noh, S. Huang, D. Leykam, Y. D. Chong, K. P. Chen, and M. C. Rechtsman, Experimental observation of optical Weyl points and Fermi arc-like surface states, Nature Physics **13**, 611 (2017).

[18] J. Y. Lin, N. C. Hu, Y. J. Chen, C. H. Lee, and X. Zhang, Line nodes, Dirac points, and Lifshitz transition in two-dimensional nonsymmorphic photonic crystals, Physical Review B **96**, 075438 (2017).

[19] Q. Guo, B. Yang, L. Xia, W. Gao, H. Liu, J. Chen, Y. Xiang, and S. Zhang, Three dimensional photonic Dirac points in metamaterials, Physical Review Letters **119**, 213901 (2017).

[20] B. Xie, H. Liu, H. Cheng, Z. Liu, S. Chen, and J. Tian, Experimental Realization of Type-II Weyl Points and Fermi Arcs in Phononic Crystal, Physical Review Letters **122**, 104302 (2019).

[21] Z. Liu, Q. Zhang, F. Qin, D. Zhang, X. Liu, and J. J. Xiao, Type-II Dirac point and extreme dispersion in one-dimensional plasmonic-dielectric crystals with off-axis propagation, Physical Review A **99**, 043828 (2019).

[22] T. Ozawa, H. M. Price, A. Amo, N. Goldman, M. Hafezi, L. Lu, M. Rechtsman, D. Schuster, J. Simon, and O. Zilberberg, Topological photonics, Reviews of Modern Physics **91**, 015006 (2019).

[23] O. Bahat-Treidel, O. Peleg, M. Grobman, N. Shapira, M. Segev, and T. Pereg-Barnea, Klein tunneling in deformed honeycomb lattices, Physical Review Letters **104**, 063901 (2010).

[24] X. Ni, D. Purtseladze, D. A. Smirnova, A. Slobozhanyuk, A. Alù, and A. B. Khanikaev, Spin-and valley-polarized one-way Klein tunneling in photonic topological insulators, Science Advances **4**, eaap8802 (2018).

[25] X. Zhang, Observing Zitterbewegung for photons near the Dirac point of a two-dimensional photonic crystal, Physical Review Letters **100**, 113903 (2008).

[26] F. Dreisow, M. Heinrich, R. Keil, A. Tünnermann, S. Nolte, S. Longhi, and A. Szameit, Classical simulation of relativistic Zitterbewegung in photonic lattices, Physical Review Letters **105**, 143902 (2010).

[27] W. Ye, Y. Liu, J. Liu, S. A. Horsley, S. Wen, and S. Zhang, Photonic Hall effect and helical Zitterbewegung in a synthetic Weyl system, Light: Science & Applications **8**, 49 (2019).

[28] L.-H. Wu and X. Hu, Scheme for achieving a topological photonic crystal by using dielectric material, Physical Review Letters **114**, 223901 (2015).

[29] X. Wu, Y. Meng, J. Tian, Y. Huang, H. Xiang, D. Han, and W. Wen, Direct observation of valley-polarized topological edge states in designer surface plasmon crystals, Nature Communications **8**, 1304 (2017).

[30] Z. Zhang, Q. Wei, Y. Cheng, T. Zhang, D. Wu, and X. Liu, Topological Creation of Acoustic Pseudospin Multipoles in a Flow-Free Symmetry-Broken Metamaterial Lattice, Physical Review Letters **118**, 084303 (2017).

[31] M. Wang, L. Ye, J. Christensen, and Z. Liu, Valley Physics in Non-Hermitian
19